\documentclass{article}
\usepackage{breckenridge}
\input psfig.sty
\frompage{000} \topage{000}                                              
\def\rightmark{Coulomb Dissociation of $^8B$}
\def\leftmark{Moshe Gai}
\title{{\bf Solar Fusion and The Coulomb Dissociation of $^8B$;}\\ 
What Have We Learned and Where Do We Go From Here?} 
\authors{
{Moshe Gai$^a$ %
}\\[2.812mm]
{\normalsize
\hspace*{-8pt} Laboratory for Nuclear Sciences, Department of Physics, U3046,
\\ 2152 Hillside Rd., University of Connecticut, Storrs, CT 06269-3046, USA \\ [0.2ex] 
}}
\abstract{The much needed nuclear input to the Standard Solar Model, $S_{17}(0)$, 
has now been measured with high precision ($\pm$5\% or better) by different groups and 
good agreement is found, even when very different methods are employed. 
We review the decade long research program to measure the cross section of the 
$^7Be(p,\gamma)^8B$ reaction using the Coulomb dissociation 
method, including the pioneering RIKEN1 experiment carried out during March 1992,
followed by RIKEN2, GSI1, GSI2 and an MSU experiment. Our RIKEN and GSI data allow 
us to rule out the much tooted large E2 contribution to the Coulomb 
dissociation of $^8B$. Specifically recent results of the MSU 
experiment are not confirmed. The GSI1 and GSI2 high precision measurements are in good 
(to perfect) agreement with the newly published high precision measurements of direct capture 
with $^7Be$ targets. From these GSI-Seattle-Weizmann high precision data we 
conclude that the astrophysical cross section factor, $S_{17}(0)$, is most likely 
in the range of 20 - 22 eV-b. We point out to an additional large uncertainty 
(-10\% +3\%) that still exists due to uncertainty in the measured slope of the S-factor and the 
theoretical extrapolation procedure which may still lower $S_{17}(0)$ down to 
approximately 18.5 eV-b. For quoting $S_{17}(0)$ with an uncertainty 
of $\pm$5\% or better, yet another measurement needs to be performed at very low energies,
as recently discussed by the UConn-Weizmann-LLN collaboration for the CERN/ISOLDE facility.}
\begin{document}
\maketitle

\setcounter{page}{1}

\section{Introduction}\label{intro}

The discovery of solar neutrino oscillations \cite{Ahm01} opens a window of opportunity 
for the study of neutrino masses with very small mass differences. While the SNO 
result \cite{Ahm01} for the measured total solar neutrino flux is rapidly approaching 
the accuracy of $\pm$5\%, the uncertainty in the Standard Solar Model is still 
dominated by nuclear inputs, with the most disturbing uncertainty in the astrophysical 
cross section factor of the $^7Be(p,\gamma)^8B$ reaction, $S_{17}(0)$ \cite{Adel}. It is 
desirable to measure $S_{17}(0)$ with accuracy comparable to the design goal 
accuracy of the SNO experiment of $\pm$5\%. For example a significantly smaller 
total neutrino flux measured by SNO as compared to an accurate prediction of the 
Standard Solar Model may teach us about oscillation of solar neutrinos into  
sterile neutrinos.

\section{The Coulomb Dissociation of $^8B$}\label{Coulomb}

The Coulomb dissociation of $^8B$ \cite{Bauer} has been suggested as a viable method 
to measure the cross section of the $^7Be(p,\gamma)^8B$ reaction. After the pioneering 
experiment of the RIKEN1 group \cite{Mot94} several experiments 
were carried out at medium energy heavy ion facilities \cite{Kik,Iw99,Dav01,Sch03} 
using a variety of kinematical regions and different experimental techniques. 
While already the data of RIKEN1 suggest a small if not negligible E2 contribution 
\cite{GaiBer} to the Coulomb dissociation of $^8B$, the MSU group claimed to have 
measured a large effect \cite{Dav01} in the measured asymmetry. The MSU model 
dependent claim has now been tested with the new GSI2 data 
\cite{Sch03}, including a test of the claim of a 
measured large asymmetry due to E2 contribution. No large E2 contribution 
was observed in the GSI2 data \cite{Sch03},
as summarized in Table 1. Note that while the RIKEN-GSI results for 
the astrophysical cross section factors were measured with increasingly higher accuracy, 
reaching the accuracy of $\pm$5\% or better, the quoted central value has risen over the 
years and stabilized around 20.5-20.8 eV-b, as can be seen in Table 1. 
The smaller value for $S_{17}(0)$ quoted by the MSU group, 
see Table 1, is almost entirely due to their model dependent assumption 
(and not a measurement) of large E2 contribution to the Coulomb dissociation 
of $^8B$.

\begin{table}[hb] 

\vspace*{-12pt}

\caption[]{Measured cross section factors in experiments on the Coulomb 
      dissociation of $^8B$. Extrapolated $S_{17}(0)$values are using 
     the theory of Descouvemont and Baye \cite{DB}.}\label{tab1}

\vspace*{-14pt}

\begin{center}

\begin{tabular}{lllll}

\hline\\[-10pt]
Experiment & $S_17(0)$ &$S_{E2}/S_{E1}$(0.6 MeV)\\ 
&(eV-b) \\
\hline\\[-10pt]
RIKEN1(94) \cite{Mot94,GaiBer} & $16.9 \pm 3.2$ & $< 7 \times 10^{-4}$ \\
RIKEN2(98) \cite{Kik} & $18.9 \pm 1.8$ & $< 4 \times 10^{-5}$ \\
GSI1(99) \cite{Iw99} & $20.6 \ + 1.2 \ - 1.0$ & $< 3 \times 10^{-5}$ \\
GSI2(03) \cite{Sch03} & $20.8 \pm 0.5 \pm 0.5$ & $< 3 \times 10^{-5}$ \ \ \ \ (Yield $< 1\%$) \\
\   \\
MSU(01) \cite{Dav01} & $17.8 \ + 1.4- 1.2$ & $(4.7\ +2.0-1.3) \times 10^{-4} $ \\
{\phantom{$00$}}\\
\hline 
\end{tabular}
\end{center}
\end{table}

\section{Comparison of Results}\label{comaprison}

\centerline{\psfig{figure=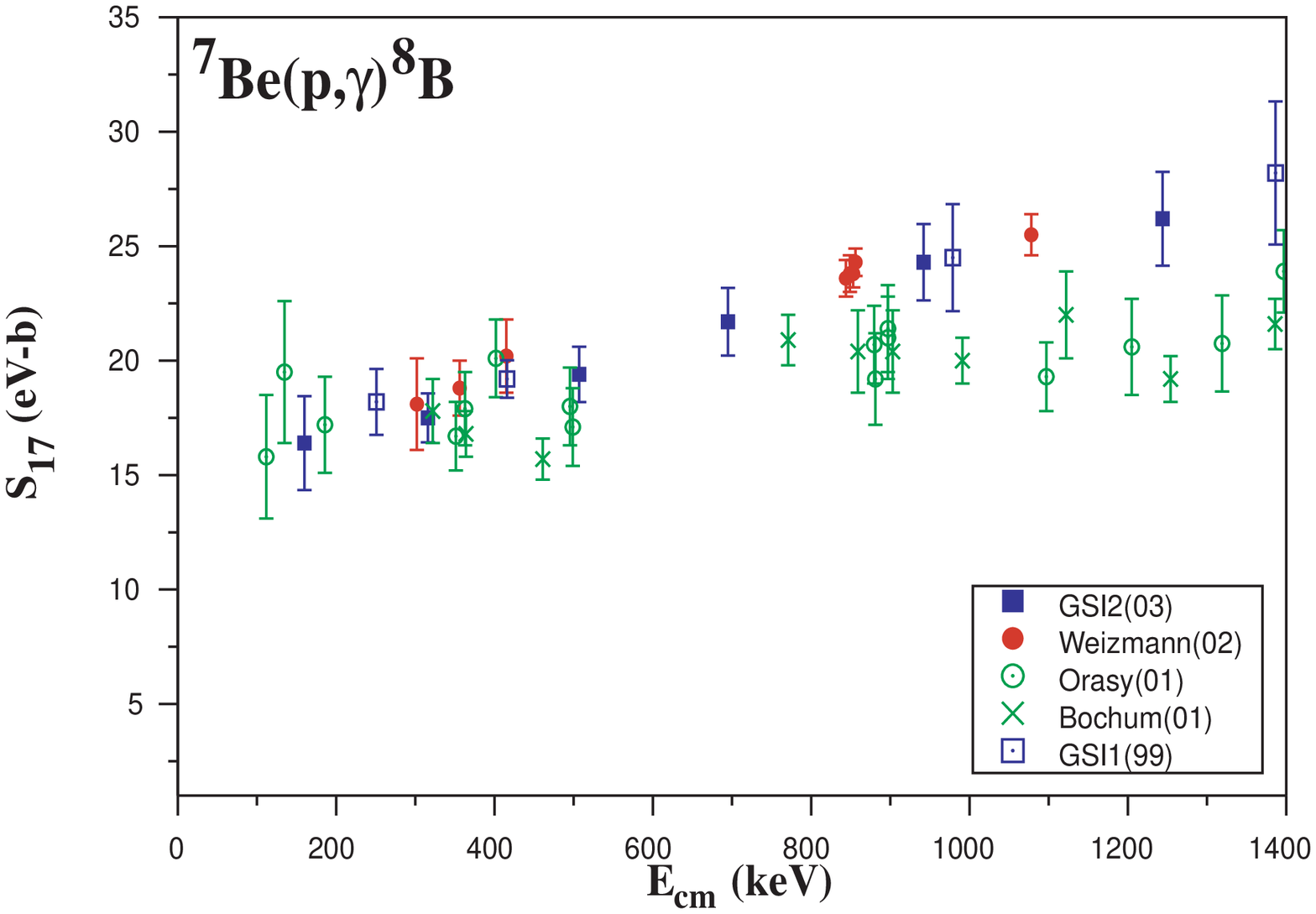,height=2in}}

\underline{Fig. 1:}  Comparison of the recent GSI2(03) 
\cite{Sch03} data, with the published data of GSI1(99) \cite{Iw99}, 
Weizmann(02) \cite{Weiz}, Orasy(01) \cite{Ham01}, and 
Bochum(01) \cite{Str01}.

%

\   \\
The recent large number of direct capture measurements of 
the $^7Be(p,\gamma)^8B$ reaction with 
$^7Be$ targets \cite{Ham01,Str01,Seatt,Weiz},  at specific energies allow us to 
perform a detailed comparison. At first we focus our attention on comparing data points 
measured at a specific energy, $S_{17}(E)$. In Figure 1 we show a comparison of 
the GSI2(03) \cite{Sch03}, GSI1(99) \cite{Iw99}, Weizmann(03) \cite{Weiz}, 
Orsay(01) \cite{Ham01}, and Bochum(01) \cite{Str01} data. The GSI1(99), GSI2(03)
and Weizmann(03) data are measured with high accuracy ($\pm5$\% or better) and 
are in very good agreement among themselves. Unfortunately this is not the case 
for the Orsay(01) and Bochum(01) data that at some energies exhibit more than 
3$\sigma$ deviation from the GSI-Weizmann data, see Figure 1.

\centerline{\psfig{figure=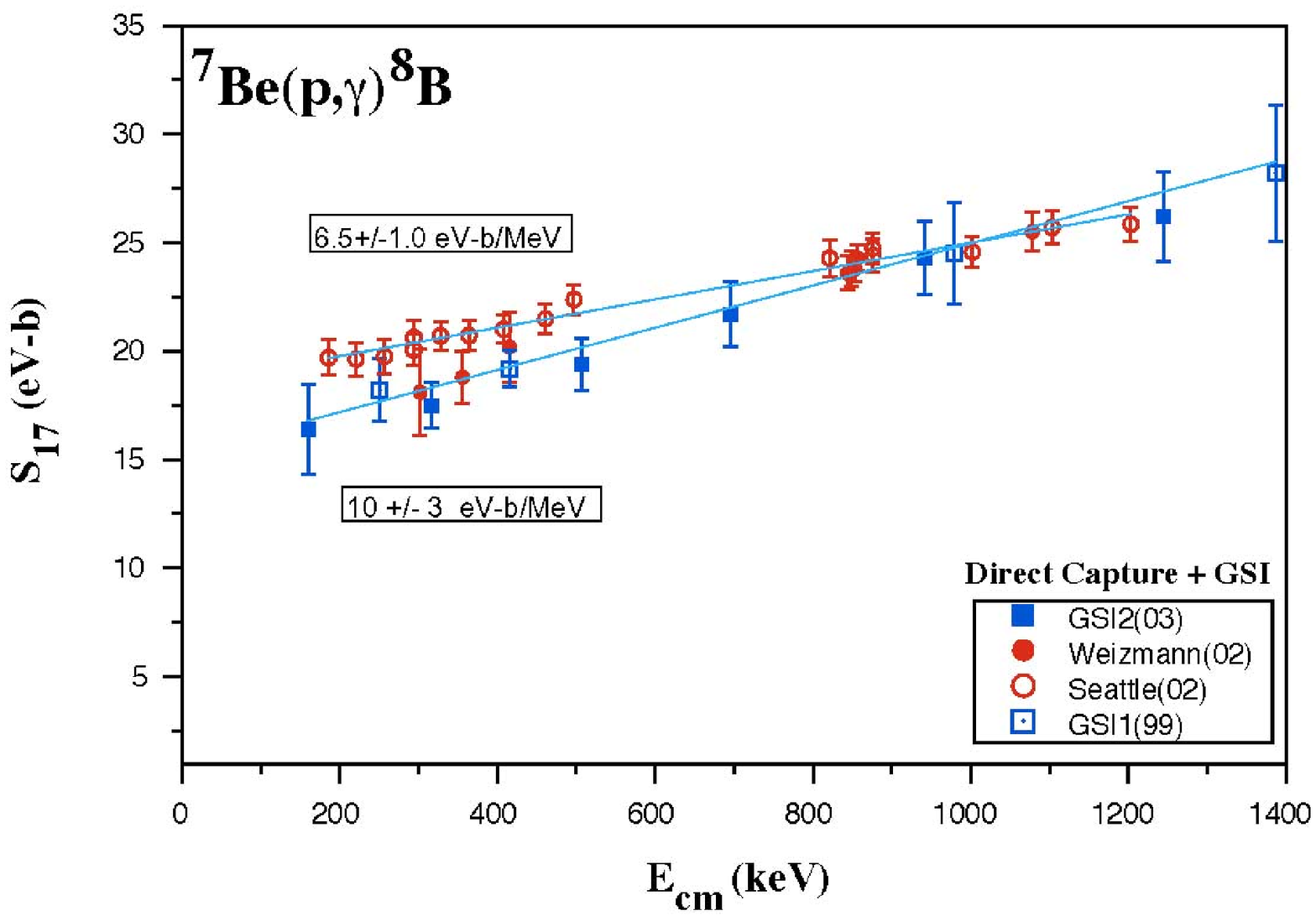,height=2in}}

\underline{Fig. 2:}  Comparison of the recent GSI2(03) \cite{Sch03} data, with the published 
data of GSI1(99) \cite{Iw99}, Weizmann(03) \cite{Weiz}, and Seattle(02)  
\cite{Seatt}. In spite of the good agreement, the measured 
slopes of the astrophysical cross section factor are sufficiently different 
that it precludes an accurate ($\pm$5\%) extrapolation to zero energy.

%

\   \\
The Seattle data \cite{Seatt} on the other hand show a remarkable agreement 
with the GSI-Weizmann data, and it is also measured with high precision. We 
conclude that the GSI-Seattle-Weizmann data could serve as a bench mark  
for studying for example the energy dependence and the slope of the measured 
s-factors. However as shown in Figure 2, while the data are in fairly good 
agreement they exhibit sufficiently different slopes, as well as sufficiently 
different absolute values at low energy, inhibiting 
an accurate ($\pm$5\%) extrapolation to zero energy.

\begin{table}[hb] 

\vspace*{-12pt}

\caption[]{Extrapolated cross section factors using the theory of 
Descouvemont and Baye \cite{DB}. Only high precision results, $S_{17}(0)$ 
measured with an error of $\pm5$\% or better, are shown, excluding the 
results of: RIKEN2(98) ($18.9 \pm 1.8$) \cite{Kik}, Orsay(01) 
($18.8 \pm 1.7$) \cite{Ham01} and Bochum(01) ($18.4 \pm 1.6$) 
\cite{Str01}.}\label{tab2}

\vspace*{-14pt}

\begin{center}

\begin{tabular}{lllll}

\hline\\[-10pt]
Experiment & $S_17(0)$\\ 
&(eV-b) \\
\hline\\[-10pt]
GSI1(99) \cite{Iw99} & $20.6 \ + 1.2 \ - 1.0$  \\
Seattle(02) \cite{Seatt} & $22.3 \pm 0.7$ \\
Weizmann(03) \cite{Weiz} & $21.2 \pm 0.7$ \\
GSI2(03) \cite{Sch03} & $20.8 \pm 0.5 \pm 0.5$  \\
\   \\
\underline{Average:} & $21.2 \pm 0.8 \ \ (\chi^2= 1.2$) \\
{\phantom{$00$}}\\
\hline 
\end{tabular}
\end{center}
\end{table}

\section{Extrapolation Methods}\label{extrapolation}

\centerline{\psfig{figure=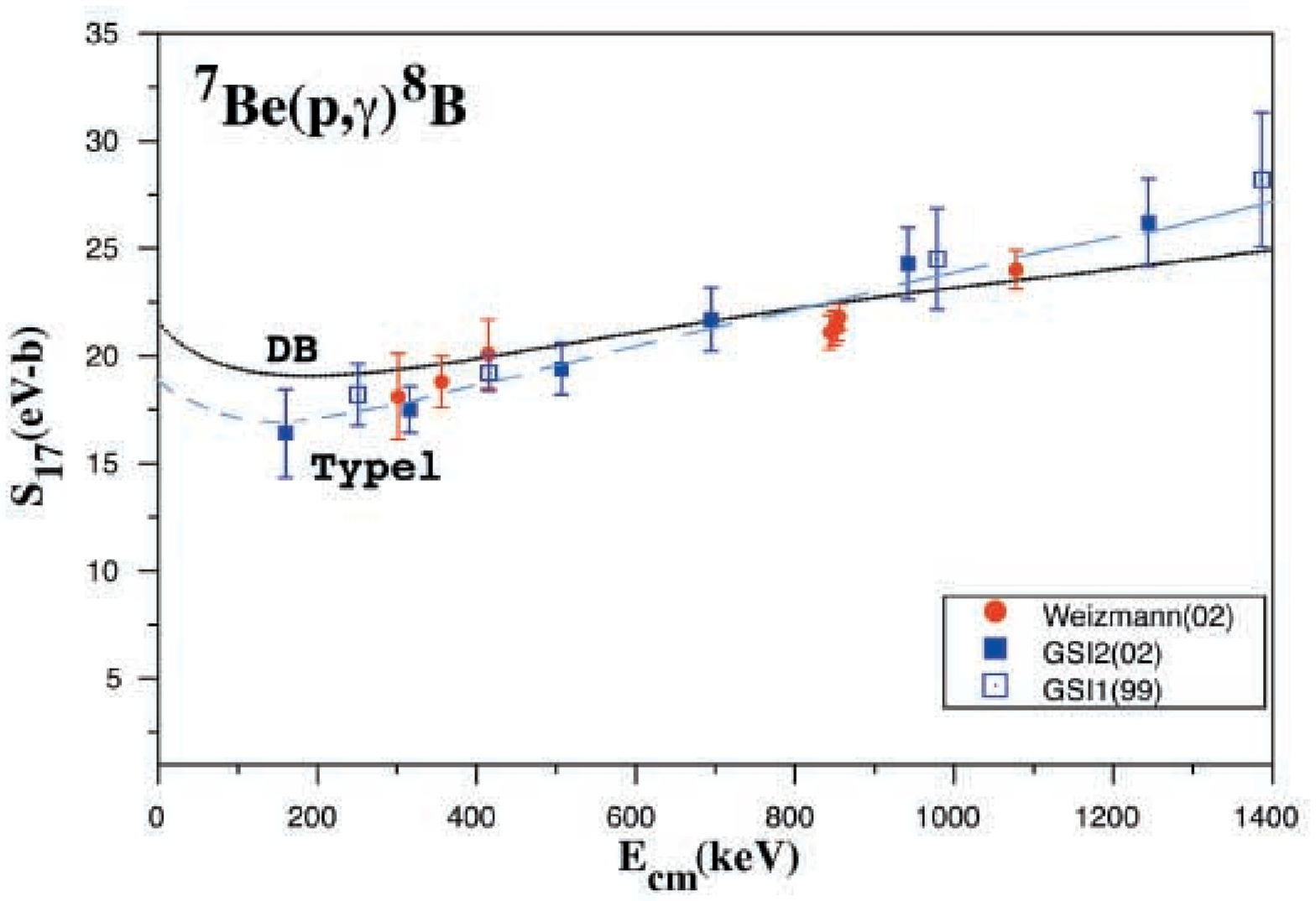,height=1.9in}}

\underline{Fig. 3:} The GSI1(99) \cite{Iw99}, GSI2(03) \cite{Sch03} and 
Weizmann(03) \cite{Weiz} data compared to the standard extrapolation 
of Descouvemont and Baye \cite{DB} and the more recent potential model 
extrapolation of Ref. \cite{Sch03}. The Weizmann(03) data point at 
approximately 850 keV is plotted with M1 contribution subtracted.

%

\  \\
In Figure 3 we show a comparison of the GSI-Weizmann data with the 
extrapolation method of Descouvemont and Baye \cite{DB} that so 
far has been used by all current experiments. The slope of the theoretical 
curve is somewhat flatter than that of the GSI-Weizmann data. The slope of 
the Seattle data on the other hand is in agreement with 
the theoretical prediction of Descouvemont and Baye. In the same 
Figure we also show a more recent theoretical curve of Typel \cite{Sch03} 
that exhibit a steeper energy dependence, and is consistent with the 
GSI-Weizmann data, but not with the Seattle data. Using Typel's extrapolation 
we deduce $S_{17}(0)$ = 18.6 $\pm$ 0.5 $\pm$ 1.0 eV-b 
\cite{Sch03}. Note that Typel model 
is a simple potential model with a variation in the potential parameter 
that yield  a different S-factor already for the s-wave component without 
altering the d-wave component. The confusion between the different 
theoretical extrapolations, in addition to the discussion in the previous 
section, does not allow us to quote $S_{17}(0)$ with the desired accuracy 
of $\pm$5\%.

\section{Future Experiment: The CERN/ISOLDE project}\label{future}

In order to resolve the issues discussed in section 4 and 3 it is very 
desirable to perform a precision measurement at low energy as we recently 
discussed at CERN/ISOLDE \cite{ISOLDE}, where they have developed the most 
intense $^7Be$ beam of up to 100 nA. In Figure 4 we show 
a possible setup discussed for this experiment where the cross section 
can be measured at $E_{cm}$ = 500-100 keV with a possible extension to 70 keV.

\centerline{\psfig{figure=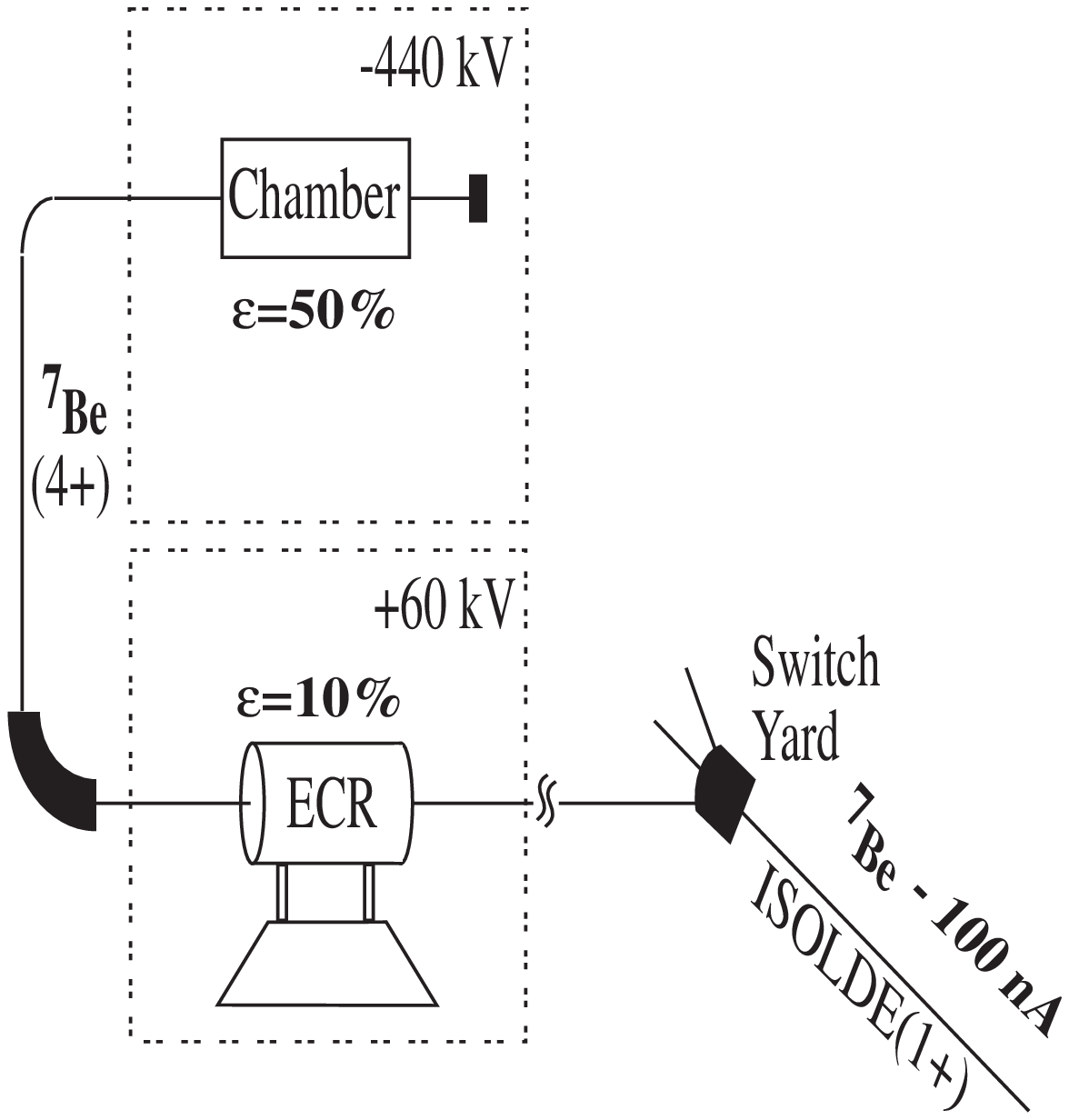,height=3in}}

\underline{Fig. 4:} The suggested arrangement of the 
 CERN/ISOLDE experiment \cite{ISOLDE}.

%

\section{Conclusions}\label{concl}

A great deal of progress has been achieved in measuring   
the cross section of the $^7Be(p,\gamma)^8B$ reaction and 
$S_{17}(E)$, that are now measured with a precision of $\pm$5\%.
But the current state of the extrapolation involving the measured 
slope as well as the theoretical model used for the extrapolation, 
still do not allow to extract the relevant nuclear 
input to the Standard Solar Model, $S_{17}(0)$, with 
the needed precision of $\pm$5\%. Instead one may conclude that the value  
is most likely in the range of 20-22 eV-b, but further study is required 
to test the possible lowering of $S_{17}(0)$ to approximately 18.5 eV-b 
due to extrapolation. A measurement of the cross section factor at low 
energies is needed to resolve the issue of the extrapolation to zero energy.

\section*{Acknowledgments}
We acknowledge discussions with Klaus Suemmerer, Stefan Typle, 
Michael Hass and Carlos A. Bertulani.
Work Supported by USDOE Grant No. DE-FG02-94ER40870.

\begin{notes}

\item[a]
Permanent address: University of Connecticut, Storrs, CT 06269-3046, USA\\ 
E-mail: gai@uconn.edu, URL: http://www.phys.uconn.edu
\end{notes}

\vfill\eject

\end{document}